InSight: AI Mobile Screening Tool for Multiple Eye Disease Detection using Multimodal Fusion

*Insight: AI for detecting multiple eye diseases*

*Ananya Raghu[1*], Anisha Raghu[1*], Alice S. Tang[4], Yannis M. Paulus[2,3], Tyson N. Kim[5**†], Tomiko T. Oskotsky[4,6**†]*

*[1]Quarry Lane School, Dublin, CA, USA*

*[2]Wilmer Eye Institute, Department of Ophthalmology, Johns Hopkins University, Baltimore, MD, USA*

*[3]Department of Biomedical Engineering, Johns Hopkins University, Baltimore, MD, USA*

*[4]Bakar Computational Health Sciences Institute, University of California San Francisco, San Francisco, CA, USA*

*[5]Department of Ophthalmology, University of California San Francisco, San Francisco, CA, USA*

*[6]Division of Clinical Informatics and Digital Transformation, University of California San Francisco, San Francisco, CA, USA*

***[*]These authors contributed equally to this work (co-first).***
***[**]These authors jointly supervised the work (co-last).***

***[†]Corresponding Authors Email Addresses:*** Tomiko.Oskotsky@ucsf.edu, Tyson.Kim@ucsf.edu

## **Structured Abstract**

### Background/Objectives

Age-related macular degeneration, glaucoma, diabetic retinopathy (DR), diabetic macular edema, and pathological myopia affect hundreds of millions of people worldwide. Early screening for these diseases is essential, yet access to medical care remains limited in low- and middle-income countries as well as in resource-limited settings. We develop InSight, an AI-based app that combines patient metadata with fundus images for accurate diagnosis of five common eye diseases to improve accessibility of screenings.

### Methods

InSight features a three-stage pipeline: real-time image quality assessment, disease diagnosis model, and a DR grading model to assess severity. Our disease diagnosis model incorporates three key innovations: (a) Multimodal fusion technique (MetaFusion) combining clinical metadata and images; (b) Pretraining method leveraging supervised and self-supervised loss functions; and (c) Multitask model to simultaneously predict 5 diseases. We make use of BRSET (lab-captured images) and mBRSET (smartphone-captured images) datasets, both of which also contain clinical metadata for model training/evaluation.

### Results

Trained on a dataset of BRSET and mBRSET images, the image quality checker achieves near-100% accuracy in filtering out low-quality fundus images. The multimodal pretrained disease diagnosis model outperforms models using only images by 6% in balanced accuracy for BRSET and 4% for mBRSET.

### Conclusions

The InSight pipeline demonstrates robustness across varied image conditions and has high diagnostic accuracy across all five diseases, generalizing to both smartphone and lab captured images. The multitask model contributes to the lightweight nature of the pipeline, making it five times computationally efficient compared to having five individual models corresponding to each disease.

## **Introduction**

Age-related macular degeneration (AMD), glaucoma, diabetic retinopathy (DR), diabetic macular edema (DME), and are leading causes of blindness. Early screening for these diseases is essential, yet access to medical care remains limited in low- and middle-income countries (LMICs). There is a significant shortage of doctors in LMICs, which contain over 90 percent of the world's blind population (1). We develop InSight, a multistage medical screening tool that leverages patient metadata and fundus images to accurately diagnose five eye-related diseases simultaneously on a mobile phone.

DR is a complication of diabetes caused by high blood sugar levels, which leads to damage and leakage of blood vessels in the eye. This leakage can cause DME, a condition where fluid accumulates in the macula, causing blurred vision and significant vision loss if left untreated. DR and DME are leading causes of blindness and growing concerns worldwide, affecting over 103 million people globally (2), (3). In addition to DR and DME, eye health can also be affected by glaucoma. Glaucoma is a disease in which the optic nerve is damaged, often caused by increased fluid build-up in the eye leading to increased intraocular pressure and asymptomatic vision loss. Currently, approximately 80 million people worldwide have glaucoma (4), and in LMICs as many as 35% of people with glaucoma are already blind as a result (5). AMD is a disorder affecting the retina, leading to the accumulation of waste products which can lead to the formation of drusen, progressive damage to the macula, and significant central vision loss. Over 200 million people worldwide are estimated to have AMD, and by 2040 this number is projected to rise to 300 million (6). Pathological myopia is a severe form of myopia that can lead to rapid vision loss and increase the risk of blindness. It affects approximately 399 million people worldwide (7). While the damage is irreversible, early detection and proper management can help slow its progression and reduce the risk of severe vision impairment, making medical screening, particularly automated screening methods, very important. Recently, smartphone fundus imaging has emerged as a cost-effective, portable, and accessible alternative for disease screenings, allowing for integration with rapid cloud based workflows for diagnosis and enabling on-phone based diagnosis of diseases (8), (9), (10), (11).

Table 1 provides a summary of the key fundus indicators for each eye disease.

| Disease | Fundus Indicators |
|---|---|
| Diabetic Retinopathy (DR) | Microaneurysms, hemorrhage, vascular changes |
| Glaucoma | Increased cup-to-disc ratio |
| Diabetic Macular Edema (DME) | Fluid buildup around the macula, hard exudates (yellowish lipids and proteins) |
| Pathological Myopia | Visible tessellations (appearance of large blood vessels at the back of the eye) |
| Age-Related Macular Degeneration (AMD) | Drusen (yellow deposits in the retina), pigmentary changes, subretinal hemorrhage |

***Table 1.** Fundus Indicators for Various Eye Diseases*

Many AI-based approaches for eye screenings have focused on single disease detection (12), (13), (14), (15), (16) including DR, AMD, pathological myopia, and glaucoma, respectively, and typically have applied deep learning transfer models for disease detection. ImageNet-based pre-training has also been used to generate disease detection models (17), (18). ImageNet based pretraining is a common practice in computer vision, but has notable limitations when applied to medical image analysis. The domain discrepancy between images from ImageNet and fundus images can limit the effectiveness of features learned from ImageNet when transferred to other tasks (19). In addition, the features learned from ImageNet images may not align well with specific characteristics of medical images, which can lead to suboptimal performance (20). RETFound (21) has revolutionized fundus-based image analysis by applying masked autoencoding (MAE) techniques to build a foundation model. However, this model is extremely computationally intensive and currently unsuitable for applications on mobile devices. Finally, many approaches for disease detection have used only images, rather than integrating multiple data sources. Combining imaging data with clinical records, genetic information, and patient demographics can provide additional informative features for the model as well as a more comprehensive understanding, potentially aiding with disease detection (22).

Herein, we report the development and validation of InSight, an AI-powered application that integrates patient metadata with fundus images to enable the simultaneous and accurate diagnosis of five common eye diseases, with the goal of improving accessibility to vision screenings.

## Methods

### Datasets

To develop InSight, we primarily make use of the **Brazilian Multilabel Ophthalmological Dataset (BRSET)** (23). This dataset contains over 16,000 fundus images from 8,524 Brazilian patients, with clinical metadata and pathological/anatomical classification parameters. Patient age, gender, diabetes diagnosis, and duration of diabetes diagnosis are the clinical metadata of interest in this study. BRSET also contains disease diagnosis information for a variety of eye diseases, including DR, glaucoma, DME, AMD, and pathological myopia. We also use the **Mobile Brazilian Multilabel Ophthalmological Dataset (mBRSET)** (24) dataset for our experiments. This dataset contains 5,164 images captured on smartphone-based retinal fundus cameras from 1,291 patients, containing labels for DR and DME with additional clinical information. In this study, the model utilized only clinical metadata features that were shared between both the BRSET and mBRSET datasets, and the mBRSET dataset is used to test the classification performance of InSight on smartphone-captured fundus images and verify its robustness across different imaging conditions.

### Data Preprocessing

For certain metadata features such as hypertension, the information was not reported separately - instead, it was stated along with other conditions in the comorbidities column of the dataset, which contains self-reported patient clinical history. Data processing was done to extract hypertension occurrence as a binary feature from the patient's clinical history.

In addition, some patients diagnosed with DR were incorrectly recorded as not having diabetes, despite DR being a complication of diabetes. Similarly, patients diagnosed with DME should, by definition, have both diabetes and DR. The dataset was processed to ensure these consistency requirements were enforced. In the BRSET dataset, 1/3 of the patients had missing age values. These were replaced using the median age of patients with reported age information.

The BRSET dataset was split into training, validation, and test sets with a ratio of 0.5:0.25:0.25. Experiments were also performed with a combined dataset of BRSET and mBRSET, where we adopt the same split ratio. An augmented test set was created by applying blur and illumination changes to the test set, doubling its size with fifty percent of the images being of low quality and fifty percent being of acceptable quality.

## InSight Pipeline

The InSight pipeline consists of three stages as shown in Figure 1a, an image quality checker, a multimodal pretrained disease diagnosis model, and a DR grading model.

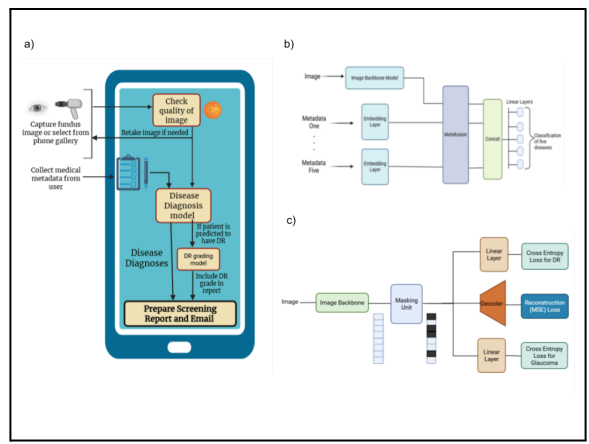

**Fig. 1. a)** *InSight app interface diagram outlining the three stage pipeline, user input, and disease diagnosis output.* **b)** *The disease diagnosis model is the core component of the InSight pipeline. It consists of a resnet18 backbone for the fundus image and embedding layers for metadata, with a MetaFusion block for multimodal fusion. The concatenated output is then passed through five linear layers to ultimately provide disease diagnoses.* **c)** *Illustration of the pre-training setup for the image backbone. We combine both self supervised and supervised pre-training, incorporating three different losses: cross entropy loss for diabetic retinopathy, cross entropy loss for glaucoma, and an image reconstruction loss.*

The pipeline follows these steps:

1. The image is captured or selected from the phone gallery.

2. The quality of the image is assessed by the image quality checker. If the image is of insufficient quality (e.g., blurriness, altered brightness), the user is prompted to retake the image.

3. When the image is of sufficient quality, it is passed to the disease diagnosis model.

4. The diagnosis for each disease is displayed to the user. If DR is determined to be the diagnosis, the stage of progression of the disease is reported.

5. Finally, a medical screening report is provided and can be emailed to the patient.

## Image Quality Checker Model

The pipeline begins with an image quality checker, which is designed to detect low-quality images that are passed as input for InSight medical screening. For training the image quality checker, we blur the images in the training set and use these along with the original images.

In order to warn users of images that are of low quality, we perform data augmentations on BRSET images to create equivalent low-quality images. The Albumentations package is used, with MotionBlur, GaussianBlur, and MedianBlur to simulate blurring due to camera motion, and Random Brightness Contrast to simulate different image exposure conditions (25). The Convolutional Neural Network (CNN) was trained as a binary classifier to classify these images as either high or low quality. The filtered images are then passed to the rest of the pipeline, to the Disease Diagnosis Model and, if appropriate, the DR Grading Model.

## Disease Diagnosis Model

The disease diagnosis model takes in fundus images and metadata and uses a fusion method to fuse images with metadata, followed by linear layers that are task specific. The model architecture is shown in Figure 1b. For efficient deployment on low-end mobile devices, we choose the ResNet18 CNN (26) as the backbone for image feature extraction. Fusion of metadata with images is done using MetaFusion (27), a fusion method that combines intermediate representations (embeddings) prior to the final layer of a model, integrating clinical metadata with imaging data. The key idea of the algorithm is to apply a correction term to image and metadata embeddings prior to concatenation using the idea of residual connections, where the correction term is based on a similarity function that measures the element-wise similarity between the embeddings.

The MetaFusion block is defined as Equation 1. Consider embeddings a and b, that correspond to different input modalities (for example $a$ could be an image embedding of dimension $d_1$ and $b$ could be an embedding for a metadata feature of dimension $d_2$). The correction to the embedding $a$ due to $b$ is given by

$$f(a,\ b) = a\ \odot\ tanh(a \odot Wb) \quad (1)$$

where W is a linear layer of size $d_1 \times d_2$ that projects $b$ into the same dimension as $a$ and $\odot$ is the element-wise product. The MetaFusion Layer consists of multiple of these correction terms (denoted by $f$) in order to modify the image and metadata, as shown in equations 2 and 3, where $x_{1,i}$ is the image embedding and $x_{2,i}$ is the embedding for the ith metadata feature, with $N$ being the total number of meta-data features. Equation 2 represents the modification of the image based on the metadata, and Equation 3 represents the modification of the metadata embeddings based on the image.

$$x_1 = x_1 + \sum_{i=1}^{N} f(x_1, x_{2,i}) \quad (2)$$

$$x_{2,i} = x_{2,i} + f\left(x_{2,i}, x_1\right) \text{ (3)}$$

All five diseases are predicted simultaneously using a single model, making it computationally efficient for mobile deployment. The loss function for training is the sum of the cross entropy losses for each of the five tasks.

To address the limited availability of data, we make use of other existing datasets of individual eye diseases. Specifically, we make use of the Eyepacs, Aptos, Messidor, Rotterdam EyePACS AIROGS, ORIGA, REFUGE, ODIR and G1020 Diabetic Retinopathy and Glaucoma datasets (28), (29), (30), (31). We combine these datasets into a larger dataset of 130000 images. To ensure that pre-training does not overfit the model to DR and Glaucoma classification, a novel loss function that combines a supervised classification loss with an unsupervised image reconstruction is used, as shown in Figure 1c.

For the DR grading model, we only consider images with DR in the BRSET dataset for training, validation, and testing. The model outputs the severity of the disease, distinguishing between mild nonproliferative DR and severe proliferative DR, thus providing information beyond just binary classification.

We evaluate the performance of the three-stage pipeline using the following metrics:

- **Balanced Accuracy (BA):** Average of true positive rate and true negative rate: used to handle imbalanced datasets. BRSET is highly imbalanced (see Supplemental Table 1).

- **Area under Receiver Operating Characteristic (AUROC):** Area under ROC curve, a graph of true positive rate vs. false positive rate at various classification thresholds.

## Results

### Overall improvements

We evaluate the end-to-end performance for disease detection on the augmented test set consisting of high and low quality fundus images. Note that only images passed through the image quality checker are considered for the evaluation of the disease diagnosis model. As shown in Table 2, our multistage pipeline consistently outperforms single-stage image-only models with an average improvement of 11.2 ± 1.8% (mean ± SD across 5-fold cross-validation) in balanced accuracy.

| Disease | Single Image Baseline: No image quality check | MetaFusion Model |
|---|---|---|
| Diabetic Retinopathy | 0.79 | 0.96 |
| Glaucoma | 0.77 | 0.85 |
| Diabetic Macular Edema | 0.83 | 0.97 |
| Pathological Myopia | 0.87 | 0.94 |
| AMD | 0.82 | 0.92 |

***Table 2.*** *Combining image quality checker with MetaFusion and pretraining gives large*

*balanced accuracy improvements.*

## Disease Diagnosis model

In this section, we analyze the performance of the disease diagnosis model in isolation, assuming that the image quality checker rejects all blurred images. We evaluate the impact of pretraining, fusion and multi-tasking.

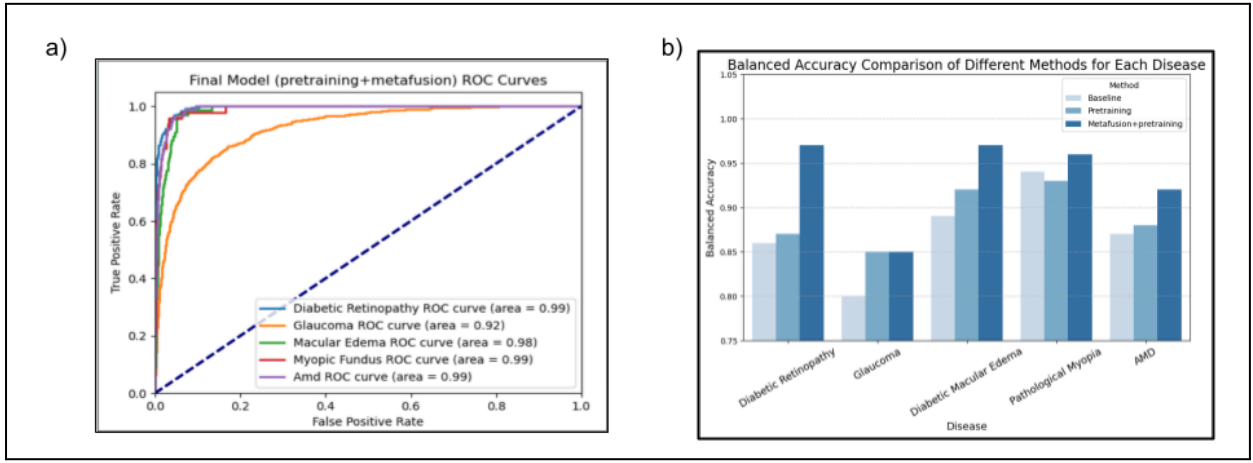

**Fig. 2. a)** *The disease diagnosis model achieves AUCs of 0.99 (DR, Pathological Myopia, AMD), 0.98 (DME), and 0.92 (Glaucoma) on the BRSET dataset. Notably, the AUC of 0.99 for DR outperforms the result of 0.97 in [9].* **b)** *Combined MetaFusion with pre training consistently has the highest balanced accuracy for all five diseases on the BRSET dataset. For DR, DME, and AMD, we observe large improvements in balanced accuracy with the addition of MetaFusion to the pretraining method. For Glaucoma, pre training significantly improves balanced accuracy, with fusion playing an insignificant role.*

We first train and test the model on BRSET. The ROC curves of the disease diagnosis model are shown in Figure 2a. The improvements in balanced accuracy due to combining pretraining and fusion are shown in Figure 2b. The pretrained multimodal disease diagnosis model shows an improvement of 6% in average BA over 5 diseases compared to a model that relies only on images. We also compare the multitask model with five individual models, each tuned for one disease in reference (see Supplemental Table 4). Due to the deployment of one model instead of five separate ones, the disease diagnosis model achieves comparable performance to individual disease-specific models at 5x lower complexity,

## Extension to Smartphone captured images (mBRSET)

To evaluate real-world performance on images taken in varied conditions, we trained our three models on a combined dataset (BRSET+mBRSET) containing both lab and smartphone camera captured fundus images.

The image quality checker trained on both datasets generalizes well on both BRSET and mBRSET, demonstrating near 100% accuracy in filtering out low-quality lab and smartphone captured images. The disease diagnosis model also generalizes well - MetaFusion and pretraining result in balanced accuracy improvements for mBRSET in addition to BRSET, improving accuracy by 4% for DR and ME, while BRSET accuracy improvements are unchanged. The improvements in accuracy of the disease diagnosis model due to MetaFusion and pretraining on the mBRSET dataset are shown in Supplemental Figure 4. To evaluate the robustness of our disease diagnosis model, we compare models trained on BRSET and mBRSET, respectively, with a single model trained on a joint dataset of BRSET and mBRSET. The model trained on the joint dataset performs as well as models trained on individual datasets, demonstrating robustness (see Supplemental Tables 5 and 6).

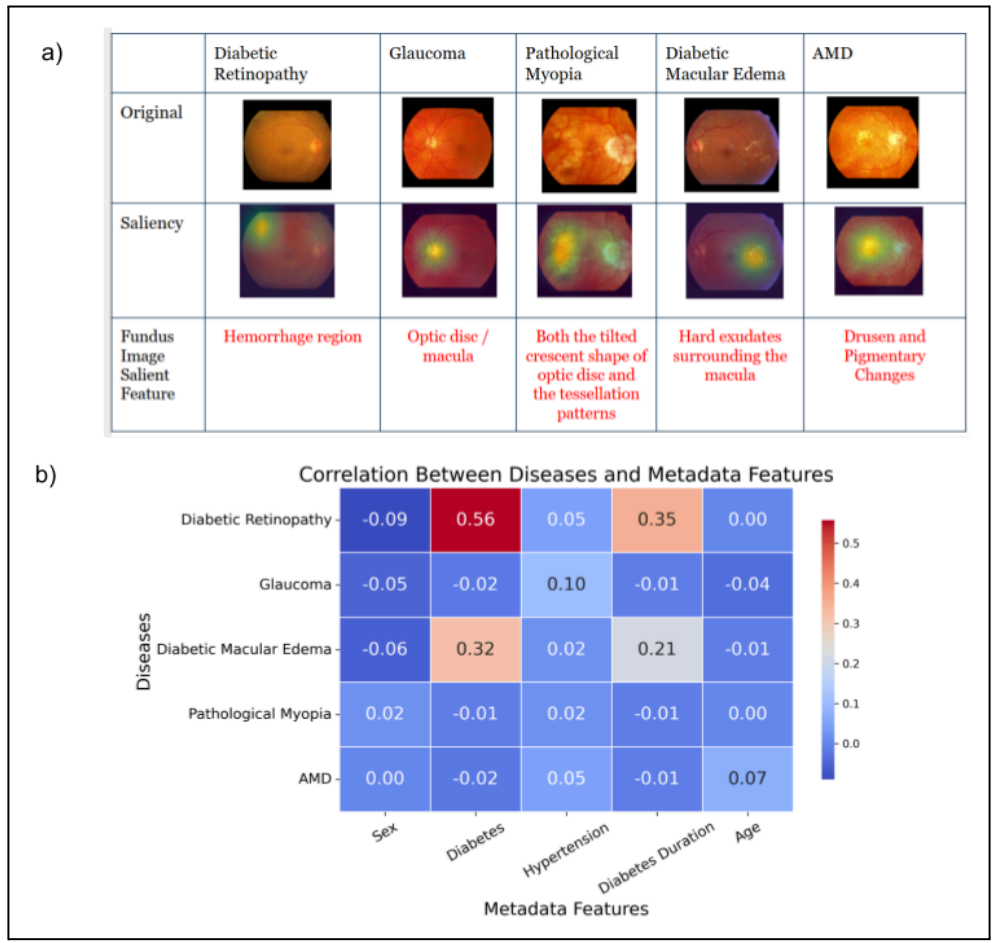

**Fig. 3.** *a)* *The model focuses on the correct fundus indicators for all diseases, providing interpretable explanations for the diagnosis, matching the fundus indicators explained in Table 1.* *b)* *The Correlation Heatmap between Diseases and Metadata Features highlights the importance of each metadata feature with each disease. We observe strong correlation between diabetes and diabetic retinopathy/diabetic macular edema.*

It is important to verify that the model uses interpretable features in the fundus image for disease diagnosis to build confidence in model predictions. As shown in Figure 3a, a saliency analysis was conducted on fundus images of each disease using a sliding window approach as previously described (32). This method works by sliding a window across the fundus image and determining the importance of a region based on the drop in accuracy when that region is blocked. The model focuses on relevant anatomical regions of the fundus image for each condition, indicating that it is learning important diagnostic information rather than relying on spurious correlations.For example, for diabetic retinopathy, the saliency map shows that the model is focusing on the upper left corner of the fundus, identifying a likely hemorrhage region.

To better understand the importance of metadata, we look at the correlation between metadata and disease labels, as shown in Figure 3b. We note that there is a high correlation between DR and DME with diabetes and duration of diabetes, as shown by the higher correlation coefficients in the heatmap. This outcome is expected, as both DR and DME are complications of diabetes and thus have a high correlation with these features. For AMD, a disease that has a high correlation with age, we observed a low correlation of 0.07, which is likely because 1/3 of the patients had their age information missing in the dataset.

We compare the impact of individual metadata features on the accuracy of diabetic retinopathy prediction in Supplemental Figure 5. We observe that when the diabetes, sex, and hypertension metadata features are added, the balanced accuracy for diabetic retinopathy prediction improves significantly. On the other hand, when age is added, there is no improvement in balanced accuracy, which is most likely due to the many patients not having reported age values in the dataset.

## Discussion

We developed InSight, a three-stage pipeline for accurate detection and diagnosis of five eye diseases consisting of an image quality checker, a disease diagnosis model, and a diabetic retinopathy grading model. The image quality checker effectively filters out low-quality images, improving the accuracy of downstream models. The multitask disease diagnosis model, operating 5x lower complexity, achieves comparable performance to individual disease-specific models, making it a lightweight and efficient solution for app deployment. Our multistage pipeline consistently outperforms single-stage image-only models with average improvements of $11.2\% \pm 1.8\%$ in balanced accuracy.

Our model achieves AUCs of 0.99 (DR, Pathological Myopia, AMD), 0.98 (DME), and 0.92 (Glaucoma) on the BRSET dataset, surpassing the previous benchmark of 0.97 using a ConvNext-V2 Large Model (12). However, the ConvNext-V2 Large model contains over 15 times more parameters, and is hence less suited for mobile deployment than the multi-task diagnosis model which is based on the Resnet18 backbone.

In addition, InSight generalizes well to both smartphone fundus camera captured images and lab captured images. On the BRSET dataset, the disease diagnosis model shows an improvement of $6.0 \pm 1.6\%$ in balanced accuracy across all five diseases compared to a model that relies only on images without pre-training enhancements. On the mBRSET dataset, there is a $4.0 \pm 1.2\%$ improvement in balanced accuracy for DR and DME. In addition, the model trained on a joint (mBRSET+BRSET) dataset performs almost as well as models trained on BRSET and mBRSET, respectively, demonstrating the robustness of the disease diagnosis model on a range of imaging conditions.

Some limitations exist to this study. Although the dataset used in our study includes 16,000 images from 8,524 patients and 5,164 smartphone images from 1,291 patients, the study population is entirely based in Brazil which may limit the generalizability of our findings. We implemented imputation techniques to mitigate the effects of missing data; however, incomplete or imputed values, particularly for key patient metadata and clinical features, may still influence classification accuracy and model robustness. Implementation of InSight with smartphone-based retinal imaging systems and screening programs in the field has yet to be studied to ensure the results are robust in a real-world screening setting. While these five diseases represent leading causes of blindness, it would be beneficial to screen for additional diseases.

## Conclusion

This work establishes a scalable and accessible foundation for AI-assisted eye disease screening by moving beyond single-disease models, leveraging multimodal data integration, and avoiding reliance on large, resource-intensive foundation models. By combining fundus imaging with clinical metadata and optimizing for deployment on lightweight platforms, this approach offers a scalable and accessible solution for screening and early diagnosis.

## Data Availability

The BRSET and mBRSET datasets were accessed through the Physionet data portal. Our code is available at https://github.com/Anisha234/InSight

## Acknowledgements

We would like to thank Professor Marina Sirota for her support in acquiring the dataset for this work.

## Conflict of Interest

The authors declare that they have no conflict of interest.

## Funding

No funding was required for this study.

## Author Contribution Statement

AR and AR were equally responsible for designing the InSight pipeline and conducting all training and testing experiments. YMP, AST, TTO, and TNK reviewed the manuscript and gave overall guidance on the ophthalmological aspects of the work. Correspondence should be directed to TTO and TNK.

## References

[1] Forrest SL, Winskell K, Bekele Y, Li S, Hann K, Soi C, et al. Does the current global health agenda lack vision?. Glob Health Sci Pract. 2023 Feb 28. https://doi.org/10.9745/GHSP-D-22-00091.

[2] Paulus YM, Gariano RF. Diabetic retinopathy: a growing concern in an aging population. Geriatrics. 2009;64:16–20.

[3] Teo ZL, Tham YC, Yu M, Chee ML, Rim TH, Cheung N, et al. Global prevalence of diabetic retinopathy and projection of burden through 2045: systematic review and meta-analysis. Ophthalmology. 2021;128(11):1580–1591. doi:10.1016/j.ophtha.2021.04.027.

[4] Davuluru SS, Frankfort BJ, Lee PP, Blumberg DM. Identifying, Understanding, and Addressing Disparities in Glaucoma Care in the United States. *Transl Vis Sci Technol*. 2023;12(10):18. doi:10.1167/tvst.12.10.18.

[5] Hu VH. The global challenge of glaucoma. *Community Eye Health*. 2021;34(112):61.

[6] Vyawahare H, Shinde P. Age-Related Macular Degeneration: Epidemiology, Pathophysiology, Diagnosis, and Treatment. *Cureus*. 2022 Sep 26;14(9):e29583. doi:10.7759/cureus.29583.

[7] Holden BA, Fricke TR, Wilson DA, Jong M, Naidoo KS, Sankaridurg P, et al. Global prevalence of myopia and high myopia and temporal trends from 2000 through 2050. Ophthalmology. 2016 May;123(5):1036–1042.

[8] Ahn SJ, Kim YH. Clinical Applications and Future Directions of Smartphone Fundus Imaging. *Diagnostics (Basel)*. 2024 Jun 30;14(13):1395. doi:10.3390/diagnostics14131395.

[9] Kim TN, Myers F, Reber C, Loury PJ, Loumou P, Webster D, Echanique C, Li P, Davila JR, Maamari RN, Switz NA, Keenan J, Woodward MA, Paulus YM, Margolis T, Fletcher DA. A smartphone-based tool for rapid, portable, and automated wide-field retinal imaging. Transl Vis Sci Technol. 2018 Oct 1; 7(5):21

[10] Patel TP, Kim TN, Yu G, Dedania VS, Lieu P, Qian CX, Besirli CG, Demirci H, Margolis T, Fletcher DA, Paulus YM. Smartphone-based, rapid, wide-field fundus photography for diagnosis of pediatric retinal diseases. Trans Vis Sci Technol. 2019 May 30; 8(3):29

[11] Patel TP, Aaberg MT, Paulus YM, Lieu P, Dedania VS, Qian CX, Besirli CG, Margolis T, Fletcher DA, Kim TN. Smartphone-based fundus photography for screening of plus-disease retinopathy of prematurity. Graefes Arch Clin Exp Ophthalmol. 2019 Nov; 257(11):2579-2585. doi: 10.1007/s00417-019-04470-4. PMID: 31501929; PMCID: PMC6824990

[12] Nakayama LF, Pereira RM, Barbosa JHA, Ramos PL, Avila MP, Oliveira MC, et al. BRSET: A Brazilian Multilabel Ophthalmological Dataset of Retina Fundus Photos. *PLOS Digit Health*. 2024 Jul 11;3(7):e0000454. doi:10.1371/journal.pdig.0000454.

[13]    Le NT, Nguyen TL, Tran NH, Pham HT, Nguyen TT, Le NQK, et al. ViT-AMD: a new deep learning model for age-related macular degeneration diagnosis from fundus images. Int J Intell Syst. 2024;2024:e3026500. doi:10.1155/2024/3026500.

[14]    Du R, Li B, Wang L, He W, Li X, Yu H, et al. Deep learning approach for automated detection of myopic maculopathy and pathologic myopia in fundus images. Ophthalmol Retina. 2021 Dec;5(12):1235–1244.

[15]    Kovalyk O, Morales-Sánchez J, Verdú-Monedero R, Sellés-Navarro I, Palazón-Cabanes A, Sancho-Gómez JL. PAPILA: dataset with fundus images and clinical data of both eyes of the same patient for glaucoma assessment. Sci Data. 2022 Jun;9(1):291.

[16]    Zhao PY, Bommakanti N, Yu G, Aaberg MT, Patel TP, Paulus YM. Deep learning for automated detection of neovascular leakage on ultra-widefield fluorescein angiography in diabetic retinopathy. Sci Rep. 2023 Jun;13(1):9165.

[17]    Zhang Z, Deng C, Paulus YM. Advances in structural and functional retinal imaging and biomarkers for early detection of diabetic retinopathy. Biomedicines. 2024 Jun;12(7):1405. doi:10.3390/biomedicines12071405.

[18]    Khalifa NEM, Loey M, Taha MHN, Mohamed HNET. Deep transfer learning models for medical diabetic retinopathy detection. Acta Inform Med. 2019 Dec;27(5):327–332.

[19] Akter N, Fletcher J, Perry S, Simunovic MP, Briggs N, Roy M. Glaucoma diagnosis using multi-feature analysis and a deep learning technique. Sci Rep. 2022 May;12(1):8064.

[20]    Alzubaidi L, Zhang J, Humaidi AJ, Al-Dujaili A, Duan Y, Al-Shamma O, et al. Novel transfer learning approach for medical imaging with limited labeled data. Cancers (Basel). 2021;13(7):1590. doi:10.3390/cancers13071590.

[21]    Zhou Y, Zhang Y, Li M, Song J, Chang C, Zhang Y, et al. A foundation model for generalizable disease detection from retinal images. Nature. 2023;622:156–163. doi:10.1038/s41586-023-06555-x.

[22] Krones F, Marikkar U, Parsons G, Szmul A, Mahdi A. Review of multimodal machine learning approaches in healthcare. *Inf Fusion*. 2024;102:102690. doi:10.1016/j.inffus.2024.102690.

[23] Nakayama LF, Santos F, Barbosa I, Pereira R, Lima R, Oliveira C, et al. BRSET: Brazilian ophthalmological dataset [dataset]. PhysioNet. 2024. Available from: https://physionet.org/content/brazilian-ophthalmological/1.0.1/

[24] Nakayama LF, Santos F, Barbosa I, Pereira R, Lima R, Oliveira C, et al. mBRSET, a mobile Brazilian retinal dataset [dataset]. PhysioNet. 2024. Available from: https://physionet.org/content/mbrset/1.0/

[25] Albumentations. Albumentations documentation [Internet]. [cited 2025 Jul 2]. Available from: https://albumentations.ai/docs/

[26] He K, Zhang X, Ren S, Sun J. Deep Residual Learning for Image Recognition. *arXiv [Preprint]*. 2015 Dec 10. Available from: https://doi.org/10.48550/arXiv.1512.03385.

[27] Raghu A, Raghu A. Metafusion: a novel method for integrating clinical metadata with imaging modalities for medical applications. In: 2025 IEEE 22nd International Symposium on Biomedical Imaging (ISBI); 2025. p. 1–5. doi:10.1109/ISBI60581.2025.10981175.

[28]    Ascanipek. EyePACS, APTOS, Messidor diabetic retinopathy dataset [dataset]. 2023. Available from: https://www.kaggle.com/datasets/ascanipek/eyepacs-aptos-messidor-diabetic-retinopathy.

[29]    de Vente C, van Kevelaer P, Chakravarty A, Dwivedi A, Giraldo D, Fu DJ, et al. AIROGS: Artificial intelligence for robust glaucoma screening challenge. IEEE Trans Med Imaging. 2023 Sep;PP(99):1. doi:10.1109/TMI.2023.3313786.

[30]    Jain A. Glaucoma datasets [dataset]. 2023. Available from: https://www.kaggle.com/datasets/arnavjain1/glaucoma-datasets.

[31]    Mvd A. Ocular disease recognition (ODIR-5K) dataset [dataset]. 2023. Available from: https://www.kaggle.com/datasets/andrewmvd/ocular-disease-recognition-odir5k.

[32]    Britefury. Image region-level saliency using VGG-16 conv-net [Internet]. 2023. Available from: https://github.com/Britefury/deep-learning-tutorial-pydata/blob/master/TUTORIAL%2003%20-%20Image%20region-level%20saliency%20using%20VGG-16%20conv-net.ipynb.

## **Supplementary Information**

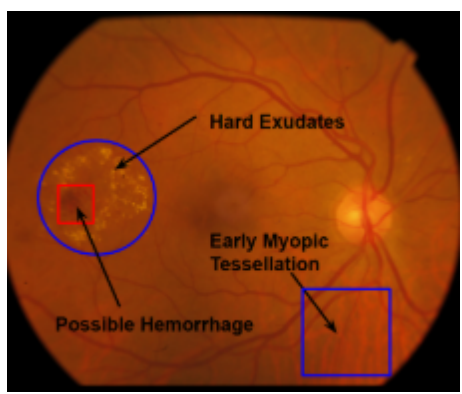

**Fig. 1**. *Annotated fundus with key anatomical structures*

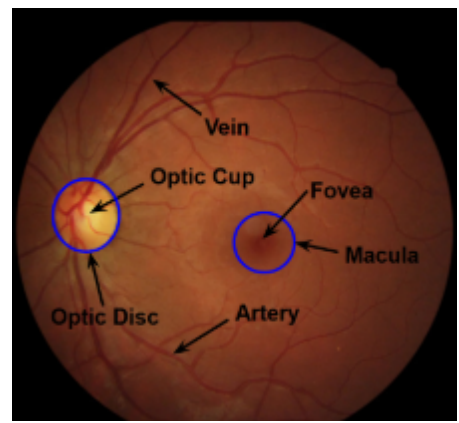

**Fig. 2**. *Annotated fundus with select abnormalities*

Key features of the fundus, showcasing both normal anatomical structures and potential disease biomarkers, are shown in Figure 1 and Figure 2 for visual reference. Figure 1 illustrates components of the healthy fundus, including the optic disc, optic cup, macula, fovea, arteries, and veins, which are important for visual function. Figure 2 highlights possible pathological abnormalities, such as hard exudates and myopic tessellation, which may indicate retinal degeneration, as well as a possible hemorrhage, a biomarker for disorders like DR. These features are essential for diagnosing and monitoring eye-related diseases. Information about the proportion of patients with each of the five diseases in the BRSET dataset is shown below in Table 1. The dataset is highly imbalanced, with a significantly higher proportion of healthy cases compared to disease cases.

| Eye Diseases | BRSET Patients With Disease (%) |
|---|---|
| Diabetic Retinopathy | 6.40% |
| Glaucoma | 19.70% |
| Diabetic Macular Edema | 2.50% |
| AMD | 2.20% |
| Pathological Myopia | 1.60% |

*Table 1.* *Statistics of BRSET dataset*

| Eye Diseases | Single Image Baseline | Single Image Baseline + Pretraining |
|---|---|---|
| Diabetic Retinopathy | 0.86 | 0.87 |
| Glaucoma | 0.86 | 0.86 |
| Diabetic Macular Edema | 0.89 | 0.92 |
| Pathological Myopia | 0.92 | 0.94 |
| AMD | 0.87 | 0.88 |

*Table 2.* *Pretraining leads to balanced accuracy improvements for all five diseases. Largest improvement was observed for Glaucoma, which makes sense as the model was pretrained extensively for Glaucoma classification. A large improvement was also observed for DME, suggesting that pretraining for the DR classification task helps with the detection of severe DR (DME is a serious complication of DR).*

| Eye Diseases | Single Image Baseline + Pretraining | Metadata Only | MetaFusion |
|---|---|---|---|
| Diabetic Retinopathy | 0.87 | 0.94 | 0.96 |
| Glaucoma | 0.86 | 0.81 | 0.85 |
| Diabetic Macular Edema | 0.92 | 0.86 | 0.97 |
| Pathological Myopia | 0.94 | 0.5 | 0.94 |
| AMD | 0.88 | 0.72 | 0.92 |

*Table 3. For certain diseases like DR and DME, the metadata-only model performs well, with balanced accuracies of 0.94 and 0.86 respectively. This is expected, as diabetes is an important predictor of both diseases. For Pathological Myopia, Glaucoma, and AMD, the single image baseline with pretraining outperforms the metadata-only model, since fundus image features provide important diagnostic information. MetaFusion combines both modalities and achieves the best performance across all diseases.*

| Eye Diseases | MetaFusion: 5 individual models for each disease | MetaFusion: 1 model predicting all diseases |
|---|---|---|
| Diabetic Retinopathy | 0.95 | 0.96 |
| Glaucoma | 0.85 | 0.85 |
| Diabetic Macular Edema | 0.95 | 0.97 |
| Pathological Myopia | 0.94 | 0.94 |
| AMD | 0.92 | 0.92 |

*Table 4. Single-task and multitask performance comparisons were also conducted with the MetaFusion method. Both demonstrated comparable balanced accuracy for all five diseases, showing how the multitask classification method achieves high performance at a much lower complexity.*

| Eye Diseases | Trained on BRSET Tested on BRSET | Trained on BRSET and mBRSET Tested on BRSET |
|---|---|---|
| Diabetic Retinopathy | 0.96 | 0.96 |
| Glaucoma | 0.97 | 0.97 |
| Diabetic Macular Edema | 0.94 | 0.93 |
| Pathological Myopia | 0.85 | 0.83 |
| AMD | 0.92 | 0.9 |

*Table 5. Evaluation on BRSET: Our disease diagnosis model generalizes well across smartphone and lab captured fundus images. A model trained on a joint (mBRSET+BRSET) dataset performs almost as well as a model trained on BRSET.*

| Eye Diseases | Trained on mBRSET Tested on mBRSET | Trained on BRSET and mBRSET Tested on mBRSET |
|---|---|---|
| Diabetic Retinopathy | 0.79 | 0.79 |
| Diabetic Macular Edema | 0.87 | 0.86 |

***Table 6.*** *Evaluation on mBRSET: Our disease diagnosis model generalizes well across smartphone and lab captured fundus images. A model trained on a joint (mBRSET+BRSET) dataset performs almost as well as a model trained on mBRSET.*

The impact of pre-training on model performance is captured in Table 2. To understand the impact of fusing metadata with images, we compare a model that uses only imaging data, a model that uses only metadata, and a model that implements MetaFusion to effectively fuse information from both modalities in Table 3. To demonstrate the low complexity of our disease diagnosis model, a side-by-side comparison of balanced accuracies for five separate models for each disease versus one model to predict all five diseases is shown in Table 4. The confusion matrix shown in Figure 3 highlights the performance of the Diabetic Retinopathy grading model. As shown in Figure 4, the bar chart shows the incremental improvement in balanced accuracy of our disease diagnosis model with the pretraining and fusion methods. To evaluate the robustness of our disease diagnosis model, we compare models trained on BRSET and mBRSET respectively with a single model trained on a joint dataset of BRSET and mBRSET. The model trained on the joint dataset performs as well as models trained on individual datasets, demonstrating robustness (Table 5 and Table 6).

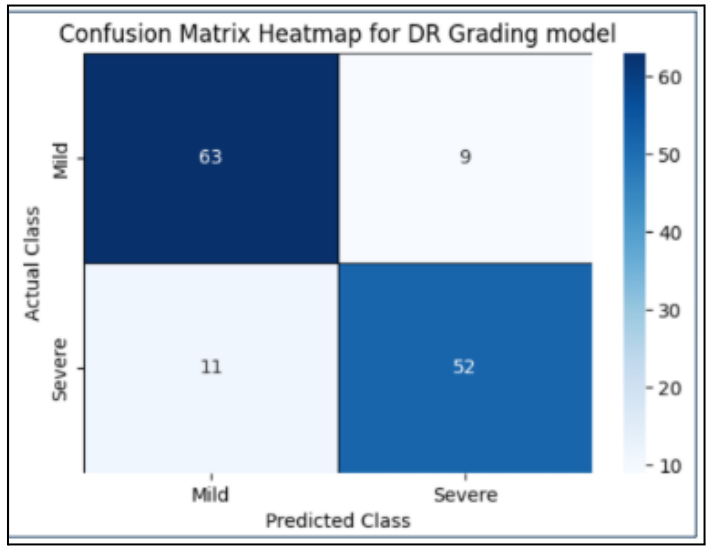

***Fig. 3***. *The DR Grading model achieves a balanced accuracy of 0.85.*

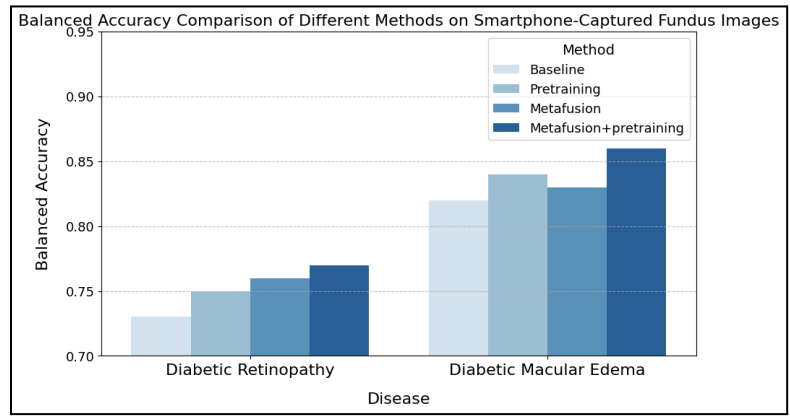

**Fig. 4.** *For a model trained on the joint dataset, combined MetaFusion and pretraining shows improvements of 4% in BA for both DR and macular edema on mBRSET.*

In Figure 5, we can observe the increasing trend in balanced accuracy with the addition of each metadata for diabetic retinopathy prediction. InSight App screens and patient reports are shown in Figure 6. The first screen shows where the patient enters in their medical history and the second screen shows fundus image capture/selections. The next few screens show disease diagnosis predictions to the patient in an interactive pie chart. The predicted diabetic retinopathy severity is shown in a text box below the pie chart. The app also provides a medical screening report with a summary of diagnoses.

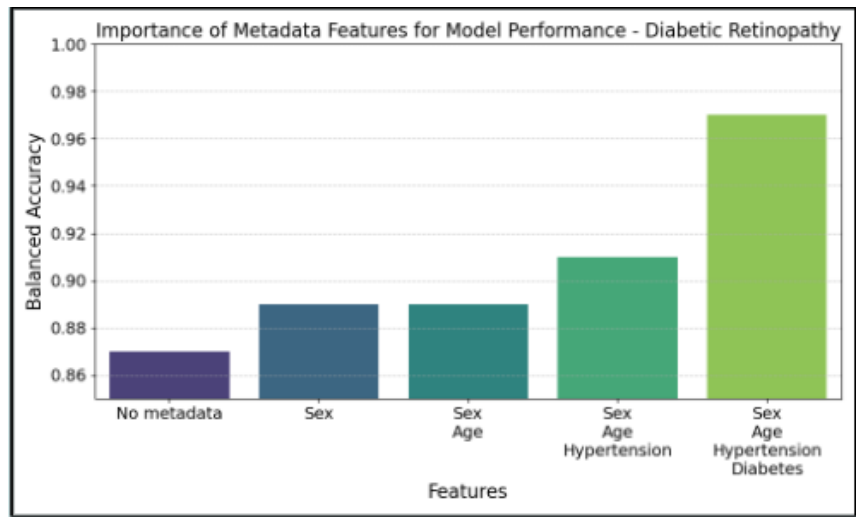

**Fig. 5.** *Here we can observe the increasing trend in balanced accuracy with the addition of each metadata for diabetic retinopathy prediction. When adding certain metadata, such as Age, we hardly see an improvement, but when Diabetes is added, there is a drastic increase in balanced accuracy. This aligns with the correlation heatmap findings in Figure 3 a) in the main paper, which highlight diabetes and diabetes duration as the most influential factors.*

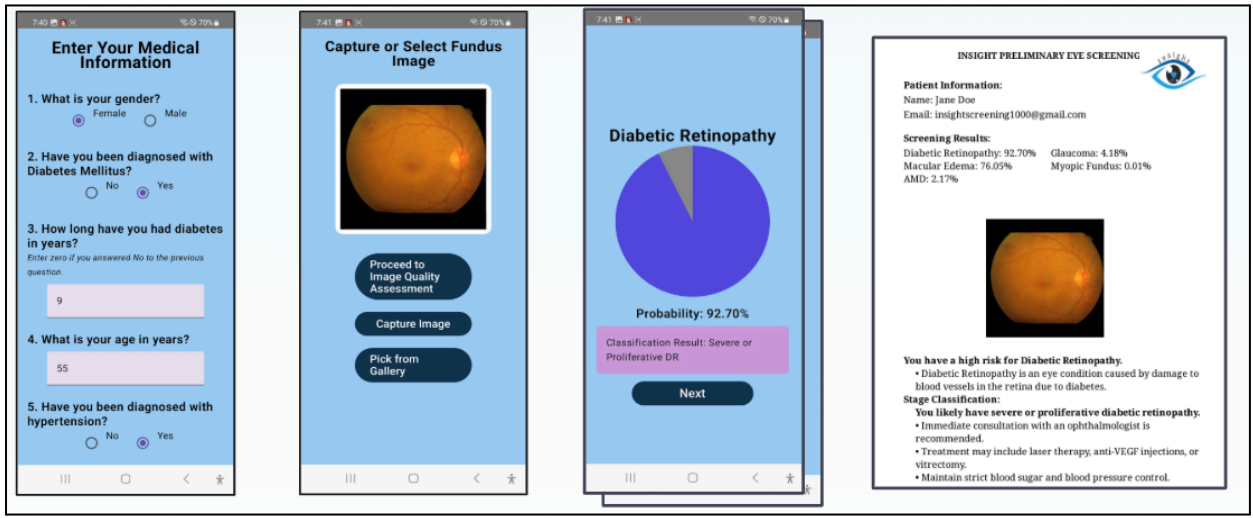

**Fig. 6.** *InSight App Screens*